# Micro-Raman spectroscopy of ultrashort laser induced microexplosion sites in silicon

L. A. Smillie,[1,2*] M. Niihori,[2] L. Rapp,[1] B. Haberl,[3] J. S. Williams,[2] J. E. Bradby,[2] C. J. Pickard,[4,5] A. V. Rode[1*]

[1]Laser Physics Centre and [2]Department of Electronic Materials Engineering,
Research School of Physics, The Australian National University, Canberra, ACT 2601, Australia

[3]Neutron Scattering Division, Neutron Sciences Directorate, Oak Ridge National Laboratory,
Oak Ridge 37831, USA

[4]Department of Materials Science and Metallurgy, University of Cambridge,
Cambridge CB3 0FS, United Kingdom

[5]Advanced Institute for Materials Research, Tohoku University,
Aoba, Sendai, 980-8577, Japan

*e-mail: lachlan.smillie@anu.edu.au; andrei.rode@anu.edu.au

Abstract

Confined microexplosions induced in silicon by powerful ultrashort laser pulses can lead to new Si phases. Some of these have not previously been observed via near-equilibrium compression of silicon. In this study, confocal Raman micro-spectroscopy and Raman imaging of arrays of microexplosions have been conducted to search for Raman signatures of these novel allotropes of silicon. A microexplosion is generated at the interface between a thick silicon dioxide confinement layer and underlying silicon. It is characterised by a void at the interface above a region of compressed silicon. Raman data show a rich assembly of silicon phases within the modified silicon. Residual stresses up to 4.5 GPa in the modifications have been determined from the shift in the main diamond-cubic Si Raman peak. The computed Raman spectra for a number of Si allotropes show reasonable agreement with the experimental spectra. Two structurally similar tetragonal phases of silicon (the rhombohedral r8 and the body-centred bc8) phases as well as recently identified bt8-Si are all highly likely to be contained in Raman spectra from many laser-modified sites. Although the st12-Si phase, previously observed in our electron diffraction studies of the highly compressively stressed laser-modified regions, was not reliably identified from Raman data, we suggest this could be due to the possible difference in residual stress level in the sites analysed by the electron diffraction and Raman spectra. Several other unidentified Raman peaks were observed, suggesting the presence of other unknown silicon phases. All of these silicon phases are expected to have attractive semiconducting properties including narrow band gap that open up novel applications.

# I. INTRODUCTION

The discovery of metastable states of matter formed by powerful ultrashort laser pulses demonstrates a new pathway for creating exotic material phases with novel optoelectronic properties [1–6]. Ultrashort laser pulses (in the sub–picosecond time domain) focused tightly inside a transparent material can deliver energy at a density of several MJ/cm$^3$ in a sub-micron confined volume. Such a level of energy density results in conversion of the laser-heated material to a high pressure/temperature solid-density plasma, or Warm Dense Matter (WDM), where the initial crystalline state is completely destroyed with no memory of the original structural arrangements [4–6]. Relaxation to ambient conditions occurs through intermediate transient states of matter resulting in novel solid metastable phases recovered from high pressure. These phases form with greatly reduced thermodynamic and kinetic constraints compared with those that apply to near-equilibrium diamond anvil cell (DAC) and nanoindentation experiments. The material is subjected to transient pressures in the order of TPas, which is higher than the strength of any material, as well as temperatures in excess of 100,000 K and ultra-high cooling rates of ~$10^{14}$ K/s. This gives access to novel, non-equilibrium material states [3,4]. These resulting metastable phases are preserved inside the surrounding pristine structure for subsequent studies and utilisation. At this level of energy concentration, new material phases with a modified energy bandgap can be formed, providing a possibility for controlled bandgap engineering with tuneable bandgap energy [6].

The ultrashort laser induced microexplosion conditions, that are very far from thermodynamic equilibrium, open up synthesis avenues to novel functional material structures which are formed at pressures well above the pressure limitation of the DAC and cannot be produced by other means. In the case of high-pressure and metastable structures of silicon (Si), the microexplosion approach offers such advantages as preservation of chemical purity of Si and precise control over the size and position of the laser-modified sites. This is unlike the approaches of increasing absorption by surface texturing [7] and by shift of the bandgap via hyperdoping in the presence of gaseous SF$_6$ [8], or by ion implantation of chalcogens followed by pulsed laser melting [9]. At the same time, laser processing can be applied to large areas by scanning a pulse train over the Si surface. We are currently at the initial stages of understanding the time-temperature-pressure conditions for the formation of the various metastable Si allotropes in non-equilibrium laser-induced confined-microexplosion experiments.

Pressure-induced phase transformations of Si and the accompanying bandgap reduction has been studied extensively under static pressure conditions, both theoretically and experimentally, because of the potential of these new materials in electronic and photovoltaic devices [10–13]. Si under increased pressure undergoes a sequence of phase transitions, some of which have been identified only recently [3,10,11,14,15]. Application of pressure to Si using nanoindentation or a DAC causes the thermodynamically stable diamond-cubic (dc, Si-I) phase to convert to the metallic *β*-Sn-Si (Si-II) phase at ~11 GPa [16]. A further increase of pressure results in a series of metal-metal transformations: it transforms via the *Imma* phase (Si-XI) into the simple hexagonal (sh) structure (Si-V) at ~16 GPa [16]; it restructures into the *Cmca* phase (Si-VI) at 38 GPa [17], into hcp (Si-VII) at 42 GPa [17], and into fcc (Si-X) at 79 GPa [18]. There is a prediction that at a much higher pressure of 250-360 GPa the fcc phase converts into a stable bcc structure [19]. This last regime has not yet been probed with DAC for the case of Si. It may be accessible via non-equilibrium conditions such as the confined microexplosion route. A peculiar property of Si is that, upon decompression, the transition to the ambient dc-Si state is not fully reversible: the *β*-Sn-Si phase transforms under static conditions into r8-Si (Si-XII), a narrow band gap semiconductor [20] at around 9 GPa and to bc8-Si (Si-III), a semimetal [21], at about 2 GPa. The bc8-Si phase is metastable at room temperature and further transforms via an intermediate transition to hexagonal diamond Si (hd-Si, Si-IV) to dc-Si on annealing at ~750°C [22–24]. However, a form of amorphous Si (a-Si) can also result during relatively rapid decreases in pressure in a DAC [25,26] or during nanoindentation [27,28], and at relatively low laser fluence (~50 J/cm$^2$) conditions in ultrashort laser induced microexplosion experiments [3]. It should be emphasised that these pathways occur for static conditions in a DAC or in nanoindentation experiments but in this work other transformation pathways to novel metastable phases under highly non-equilibrium conditions may be of greater relevance.

Recently, in the far from equilibrium microexplosion regime, transmission electron microscopy (TEM) revealed two novel phases [3], tetragonal bt8-Si (structurally an intermediate state between *β*-Sn-Si and bc8-Si with an I4$_1$/a space group [29] and similarities to both bc8-Si and r8-Si [30]) and st12-Si (P4$_3$2$_1$2 space group) which had been previously predicted in Si due to its existence within the very similar Ge system [20,31]. Furthermore, additional Si structures were observed but not characterised, noting that four additional phases, two monoclinic m32 and m*32 and two tetragonal t32 and t32* have been predicted (with quasi-direct bandgaps) through the *ab initio* random structure searching (AIRSS) approach [32,33] using the CASTEP code [33,34]. To date it has only been possible to synthesize small material quantities in microexplosion experiments. Furthermore, nanocrystals of these new phases, of ~10 – 100 nm in size, have been observed in a mixture of other Si phases [3], making detailed characterisation of fundamental physical properties of these new structures extremely challenging.

In this paper Raman micro-spectroscopy is conducted across arrays of microexplosion sites in search of possible Raman signatures of new metastable Si phases. We present Raman spectra from modifications created under a range of laser energies and discuss possible candidate phases for the observed Raman features. Alongside strong signals from compressed dc-Si and a-Si, we observe peaks that appear to correlate with a number of known phases including r8-Si, bc8-Si, bt8-Si, peaks that are suggestive of twin-like stacking, and also peaks which do not match any well described Si phases, potentially including st12-Si.

## II. EXPERIMENTAL

### A. Ultrashort laser induced confined microexplosion

In order to confine the laser–matter interaction mode inside the material bulk, avoid ablation and ensure the generation of a strong shock wave with maximum pressure above the Young's modulus, the laser radiation should be focused well below the surface. At the same time, the laser fluence at the surface of the material should be below the damage threshold. A tightly-focused interaction mode with a high-numerical aperture (NA) microscope objective is the way to satisfy these two conditions: it provides high energy density confined inside the bulk of a cold and dense solid. The laser-modified material remains in the focal area and is preserved by the high Young's modulus of the surrounding pristine material.

Si is not transparent at 790 nm, the laser wavelength that we use in our experiments. To provide the confinement conditions, we thus used Si wafers with an amorphous oxide layer on the surface. A 10-μm thick layer of silicon dioxide (SiO$_2$) was grown thermally via wet oxidation on the surface of 500-μm thick boron-doped (100) Si wafers (*p*-type Si, resistivity 1-10 Ω-cm, impurity concentration $10^{15} - 10^{16}$ cm$^{-3}$, Silicon Quest International, Inc). SiO$_2$ is transparent to the laser radiation and allows for tightly focused femtosecond laser pulses applied to the buried surface of the Si crystal. The thickness of the SiO$_2$ layer was not so deep as to develop large spherical aberrations with high-NA focusing optics but at the same time did guarantee the absence of optical breakdown and damage on the surface. The boundary between the transparent SiO$_2$ layer and crystalline Si is very sharp and can be clearly seen in the cross-section of the wafer using scanning electron microscopy (SEM) [35]. Most of the ultrashort laser pulse energy is deposited into Si (Si-plasma), as the ionisation threshold in Si is about an order of magnitude lower than in SiO$_2$, while the SiO$_2$ layer is responsible for the confinement of the deposited energy [36].

The microexplosion experiments have been conducted using 170-fs, 790-nm laser pulses from a Ti-sapphire MXR-2001 CLARK laser system. Pulses with up to 2.5 µJ per pulse were focused using an inverted optical microscope (Olympus IX70) equipped with an oil-immersion 150× objective (Olympus UAPON150XOTIRF, NA = 1.45). The focal spots have been measured using a knife-edge technique with a sharp edge of a Si (100) wafer etched at 54.74° to the surface along the (111) direction and mounted on a piezo nano-positioning stage with a 200 µm range of x-y motion (MCL, Nano-T Series). The measured focal spots have a diameter of 0.74 µm at full width at half-maximum (FWHM) level. The microexplosion experiments have been conducted with a laser pulse train with a 1 kHz repetition rate in a sample moved at a rate of 2 mm/s to guarantee a single shot per spot regime, forming well separated microexplosion sites located 2 µm apart (Fig. 1). Each of the regions was irradiated by a single laser pulse with an energy in the range from 150 nJ to 700 nJ, which roughly corresponds to a fluence range from 35 J/cm$^2$ to 150 J/cm$^2$ on the Si surface. Below this energy range, the microexplosion conditions were not sufficient to produce a strong shock wave with a pressure above the Young's modulus of SiO$_2$ and Si (~75 GPa and ~165 GPa, respectively [37]) and thus to form a void surrounded by compressed material. The laser-modified sites can be clearly optically detected within this laser fluence range. The appearance of cracks between the sites has been observed with increasing fluence, becoming common above 700 nJ. Thus, this fluence level determined the practical upper limit for the deposited laser energy in our experiments.

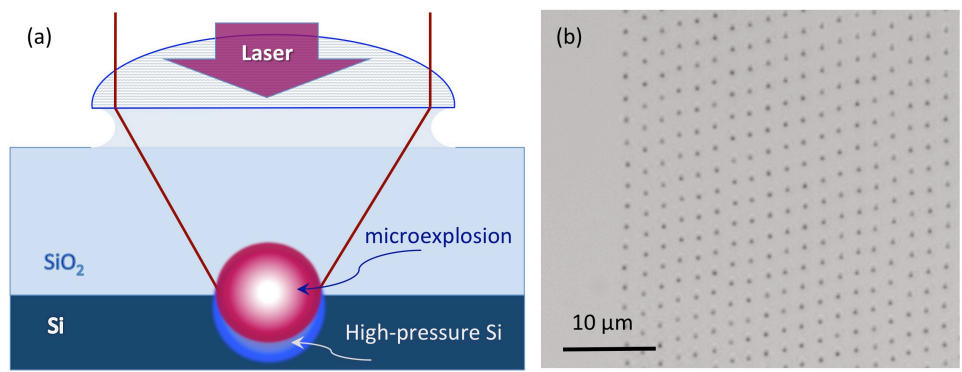

*Fig.1. (a) – Focusing conditions on the Si-crystal surface with a thermally grown SiO$_2$ layer in confined microexplosion experiments. (b) – Optical microscope image of the array of well separated microexplosion sites 2-µm apart produced by fs-laser pulses focused at the Si/SiO$_2$ interface as viewed through the oxide layer.*

B. Raman micro-spectroscopy

The Si modifications were characterised by Raman micro-spectroscopy using a Renishaw InVia Reflex Raman spectrometer. Spectra were obtained with a 532 nm laser focused on the plane of the modifications at the Si-SiO$_2$ interface using a ×100, 0.85-NA objective lens (Leica NPLAN EPI 11566073). The focal spot size was 0.75 μm and the depth of focus ~0.3 μm, which is comparable to the modification size based on previous TEM examination [3,35]. The SiO$_2$ had no observable influence on the Raman spectra collected. A 2400 lines/mm grating was used for all measurements and the spectra were collected on a Peltier cooled CCD detector. Calibration was performed on pristine dc-Si such that the transverse optical (TO) phonon band was centred at 520.5 cm$^{-1}$. The relative precision of locating peaks is within 0.1-0.5 cm$^{-1}$, where the variation is dependent on the peak strength relative to any overlapping peaks. However, the variation in peak position between modifications may be as much as 5 cm$^{-1}$ due the local morphology, particularly residual strain and grain size effects [38–41]. The spatial resolution and selectivity of the Raman spectrometer is below that of many of the structural features in the modifications. Indeed, it was previously noted that the crystalline regions of new Si allotropes are often of ~10 – 100 nm in size and this factor, coupled with their often small phase fraction and unknown, potentially limited Raman activity in comparison with more dominant Si phases, can make them difficult to detect. As such, the new Si allotropes may be visible only when and if they are present in a sufficiently large fraction of the sampled material.

Samples were initially surveyed by Raman mapping over a collection area of approximately 30×30 μm$^2$ with a 0.3-0.4 μm step size between measurements. Mapping allows for the identification of key spectral components and the visualisation of their distribution using Raman images. These Raman images are constructed by taking the area under any peak of interest, normalised against the area under the entire spectrum, for each point within the map. These Raman maps were collected with a laser power of 0.6 mW which was sufficient to avoid significant heating of the samples and hence avoided any unwanted thermal annealing of Si phases present. Some features observed during Raman mapping were re-examined in confocal conditions (narrower slit width and CCD readout). Here the laser power was reduced to 0.12 mW to prevent even subtle Raman band shifts due to heating. A linear baseline subtraction was performed before curve fitting with mixed Gaussian-Lorentzian curves using the Wires 4 software package.

III. ANALYSIS OF EXPERIMENTAL RESULTS

A. Raman mapping of microexplosion sites

Typical results of Raman mapping of microexplosion sites are presented in Figs. 2 and 3, together with the corresponding Raman spectra (Fig. 2(d) and (e)). The most intense components of the Raman spectra were the lines from dc-Si and a-Si, noting that dc-Si is the equilibrium phase of Si at ambient conditions and the initial state of the material. Within the modifications the strength of the dc-Si signal decreased, while the strength of the a-Si signal increased. This is demonstrated in Fig. 2, where (a) is an optical microscopy image of the sample irradiated by 320 nJ laser pulses. Each modification is visible with a dark dot at the centre, corresponding to the void, with a ring of compressed material around it [3]. Figure 2(b) is the Raman image (map) of the dc-Si signal from the same location. Here, brighter regions are more strongly dc-Si, although it should be emphasised that the dc-Si signal is never absent. In Fig. 2(c), the Raman map of the a-Si signal from the same location is shown. Brighter regions have a stronger a-Si signal and these clearly correspond to the location of modifications.

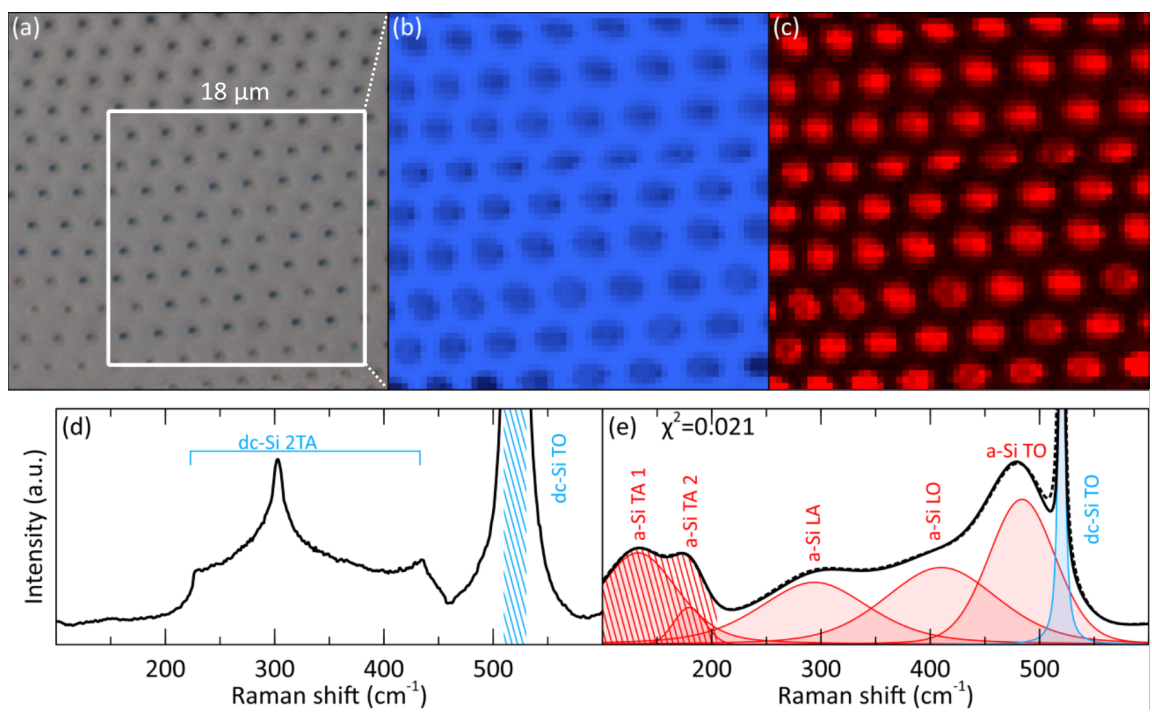

*Fig. 2. (a) optical and (b,c) Raman images of modifications induced by 320 nJ laser pulses at the Si-SiO$_2$ interface with the modification sites ~2 µm apart. (b) Raman map of the dc-Si distribution taken in the 510-530 cm$^{-3}$ range and (c) a-Si distribution (100-205 cm$^{-1}$) where the signal is stronger in the bright areas of the image; (d) labelled example spectrum from unmodified dc-Si and (e) a spectrum from a modification which also contains a-Si. Peaks are labelled and the area used to construct the Raman maps in (b) and (c) is hashed in (d) and (e). In (e) the spectrum is fitted using the Wires 4 package and the quality of the fit is expressed with the $\chi^2$ value, with the overall curve fit given by a dotted line, and the individual peaks in the fit are colour coded: blue, dc-Si and red, a-Si.*

The dc-Si Raman map was constructed by taking the area under the transverse optical (TO) phonon band of dc-Si, between 510 and 530 cm$^{-1}$ for each point in the Raman map. This band is shown but truncated in Fig. 2(d) such that the transverse acoustic (2TA) phonon band is also visible. The a-Si Raman image was constructed similarly, with the exception that the area under the TA band of a-Si was used, between 100 and 205 cm$^{-1}$. This is shown in Fig. 2(e), as is the longitudinal acoustic (LA), the longitudinal optical (LO) and the TO phonon bands of a-Si. In both cases the resultant Raman images are minimally influenced by other components within the sample and thus provide an excellent qualitative image of the spatial distribution of these phase components.

It is important to consider how dc-Si and a-Si can be curve fitted since these components must be accounted for when identifying other Si phases with much smaller phase fractions present in the spectra. For dc-Si, the 2TA band is often weak enough to be ignored within modifications. When this is not the case, three curves are fitted. Due to the unusual shape of the 2TA band, this fit is inherently poor. However, the breadth of these peaks ensures that curve fitting of any other peaks in the spectrum is largely insensitive to the relatively low quality of the dc-Si 2TA fit. For a-Si, the unusual shape of the TA band requires two curves for an adequate fit. Similar to the 2TA dc-Si band, the a-Si bands are wide enough that curve fitting of other peaks is largely insensitive to the quality of the a-Si fit.

The Raman maps can also be used to examine the effect of laser energy on the resultant modifications. Fig. 3 consists of a-Si Raman maps from a 190 nJ laser pulse energy sample (a), a 420 nJ sample (b) and a 570 nJ sample (c). In the 190 nJ sample, the a-Si appears only as spots, in the 420 nJ case rings of a-Si are resolved while in the 570 nJ sample the rings are large enough that they are no longer well separated. Simultaneously, the average intensity of the a-Si signal is decreasing, resulting in the poorly defined amorphous regions in the 570 nJ sample. It is noteworthy that this sample exhibits more variability across different modifications and significant cracking can be observed between some of the modification sites when examined by optical microscopy.

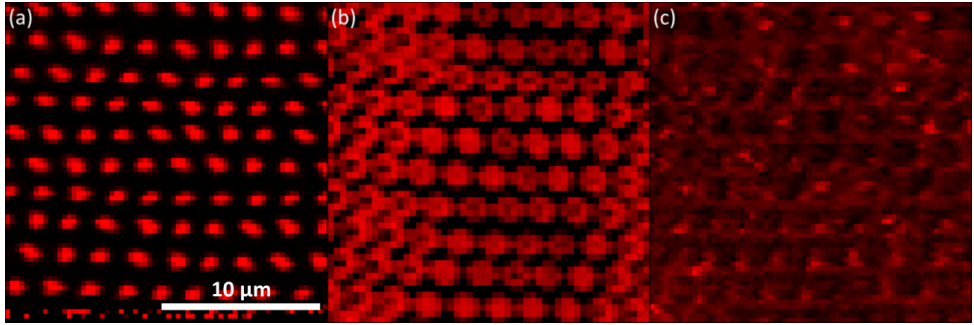

*Fig. 3. Raman images, all at the same scale, of the a-Si distribution with a laser fluence of 190 nJ (a), 420 nJ (b) and 570 nJ (c).*

B. Estimation of residual stress

In addition to the principal dc-Si and a-Si components, which were present in the Raman spectra collected from all laser-modified sites, there were a number of spectral features that were also observed in some but not all sites – see Fig. 4(a) and Fig. 5(a). These include peaks that can be associated with compressed dc-Si and several crystalline allotropes of Si, which we discuss in the following section. The laser-modified sites with these features were identified from Raman maps and re-examined under confocal Raman conditions. The results presented here were collected from samples prepared with 310 nJ or 420 nJ laser pulse energies, but it must be emphasised that these features could be observed in samples prepared across a range of laser energies, excluding only the lowest energy sample of 190 nJ where only a-Si was observed.

Several additional peaks beyond those of a-Si and dc-Si are evident in Fig. 4(a). Some of these, namely, at 522 cm$^{-1}$ and 539 cm$^{-1}$, can be explained by compressive strain, which is known to cause the TO band of dc-Si to shift to higher wavenumbers [39,40]. Furthermore, this shift in peak position can be related to compressive strain using Eq.(1) [39]:

$$\Delta\omega_{dc-Si}^{TO} = 518.6 + 0.55P - 8.66\times10^{-4}P^2 ; \tag{1}$$

where $\Delta\omega^{TO}_{dc-Si}$ is the Raman shift in [cm$^{-1}$] and *P* is the pressure in [kbar]. Inhomogeneity of strain throughout the sampled volume required three separate curves to more accurately fit the TO band. These are centred at unstrained 520 cm$^{-1}$, and strained 522 cm$^{-1}$ and 539 cm$^{-1}$ values that correspond in the latter cases to a ~0.7 GPa and ~4 GPa compressive stress, respectively. Furthermore, both strained peaks are significantly broadened relative to the unstrained peak, further indicating inhomogeneity. Using Eq.(1), residual compressive stresses of over 4.5 GPa have been estimated from the dc-Si TO peak in this study.

### C. Laser induced crystalline phases

In addition to dc-Si and a-Si peaks, the Raman spectra also contain several other peaks, such as those labelled P1 through P6 in the spectrum shown in Fig. 4(a). Another experimental Raman spectrum is shown in Fig. 5(a), which contains a further set of peaks that have been observed in multiple instances in this work but that do not match any previous Raman characterised allotropes of Si. The labelled peak positions, widths and intensities are given in Table 1 below.

*Table I. Raman curve fit parameters for experimental results presented in Fig. 4(a) and Fig. 5(a) for the various peaks corresponding to laser induced Si phases. Peak intensities are expressed relative to the most intense peaks of each case, located at 357 cm$^{-1}$ (P3) for Fig. 4(a) and 443 cm$^{-1}$ (Q4), for Fig. 5(a).*

|         | Peaks | Position [cm$^{-1}$] | Peak width [cm$^{-1}$] | Relative Intensity |
|---------|-------|----------------------|------------------------|--------------------|
| Fig. 4a | P1    | 164                  | 2.8                    | 0.47               |
|         | P2    | 169                  | 0.8                    | 0.29               |
|         | P3    | 357                  | 7.4                    | 1.00               |
|         | P4    | 388                  | 6.4                    | 0.34               |
|         | P5    | 401                  | 8.1                    | 0.32               |
|         | P6    | 447                  | 7.9                    | 0.71               |
| Fig. 5a | Q1    | 162                  | 6.1                    | 0.24               |
|         | Q2    | 393                  | 5.5                    | 0.87               |
|         | Q3    | 415                  | 16.9                   | 0.41               |
|         | Q4    | 443                  | 11.4                   | 1.00               |
|         | Q5    | 490                  | 23.8                   | 0.16               |

In order to identify reliably the observed Raman peaks, both their positions and intensities have been computed using density functional theory (DFT). Raman spectra for the dc, hexagonal diamond (hd), bc8, r8, st12 and bt8 phases of Si were computed using version 19.1 of the CASTEP code [34,42]. The structures were relaxed under zero applied external pressure and the PBEsol density functional [43]. The default OTFG norm-conserving pseudopotential was used, with a plane wave cutoff of 348.3 eV and finite basis set corrections [44]. Density functional theory is well known to underestimate electronic band gaps, and, for narrow gap systems, can predict a metallic state. The metastable phases of Si are just such narrow band gap materials, and this presents a challenge to the computation of the Raman spectra in that the intensities cannot be computed for metallic systems. In an attempt to mitigate this aspect, we do not allow partial occupancy of the electronic energy levels, and present our results for a range of Brillouin zone sampling densities: $k = 0.03\times2\pi/\text{Å}$, $0.05\times2\pi/\text{Å}$ and $0.07\times2\pi/\text{Å}$. Comparing the computed spectra for these different sampling densities allows us to assess the impact of metallisation due to the limitations of DFT. The peak positions are well described in all cases, but the Raman intensities are sensitive to the quality of the sampling, in particular for the bc8 and st12 structures, and the peak around 330 cm$^{-1}$ for the r8 structure (see Fig. 4b). The predicted spectra are presented as broadened by a Gaussian with a width of 10cm$^{-1}$ in the Figs. 4(b,c,d) for r8, bc8, and bt8 respectively, and in Fig. 5(b) for st12. Although the peak positions in the simulated spectra are in a similar wavenumber range to those in the experimental spectra, the match is not perfect for a number of reasons which we discuss below.

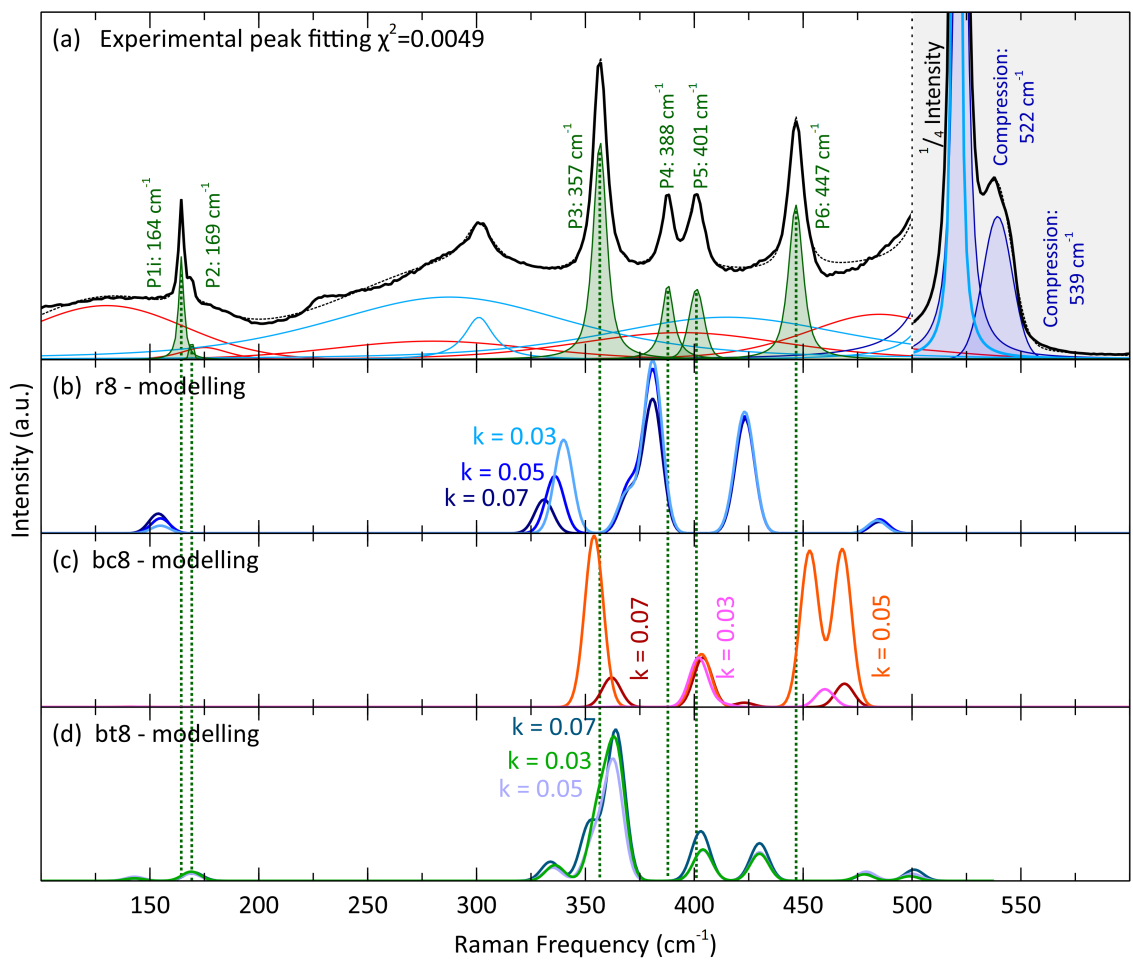

*Fig. 4. Example of a Raman spectrum collected under confocal conditions from modifications created (a) with a 420 nJ laser pulse, which contains, in addition to a-Si (see red curve fits) and dc-Si (see blue curve fits), peaks from: compressed dc-Si and additional crystalline metastable phases (labelled as P1 through P6). The peak positions in (a) are given in Table 1. Also shown are the results for computed Raman spectra in (b) r8-Si, (c) bc8-Si and (d) bt8-Si.*

As indicated, the agreement of the computed spectra with the experimentally measured Raman spectra shown in Fig. 4 is not particularly good, and though the experimental and calculated Raman spectra in Fig. 5 have similarity in general, they do not correlate in details. However, DFT gives only approximate peak positions, and around a 10% shift in the absolute peak positions is to be expected, particularly for narrow bandgap phases. Furthermore, the spectra have been computed for structures relaxed at 0 GPa, but it is quite possible that the phases observed are under some residual pressure, given that significant compressive strain was measured in terms of the shift in the dc-Si peak positions, as presented in Section III-b earlier. This would lead to further shifts in the peak positions and changes in the Raman intensities for computed spectra. There may be further phases of Si, such as the t32 and m32 phases [3], which should also be considered as having possible contributions to the experimental Raman spectra in Figs. 4(a) and 5(a). As a result, it is difficult to unequivocally assign the peaks in Figs. 4(a) and 5(a) to any of the phases calculated here, unless (in the case of Fig. 4) peak shifts with residual pressure are accounted for, as will be discussed in detail later.

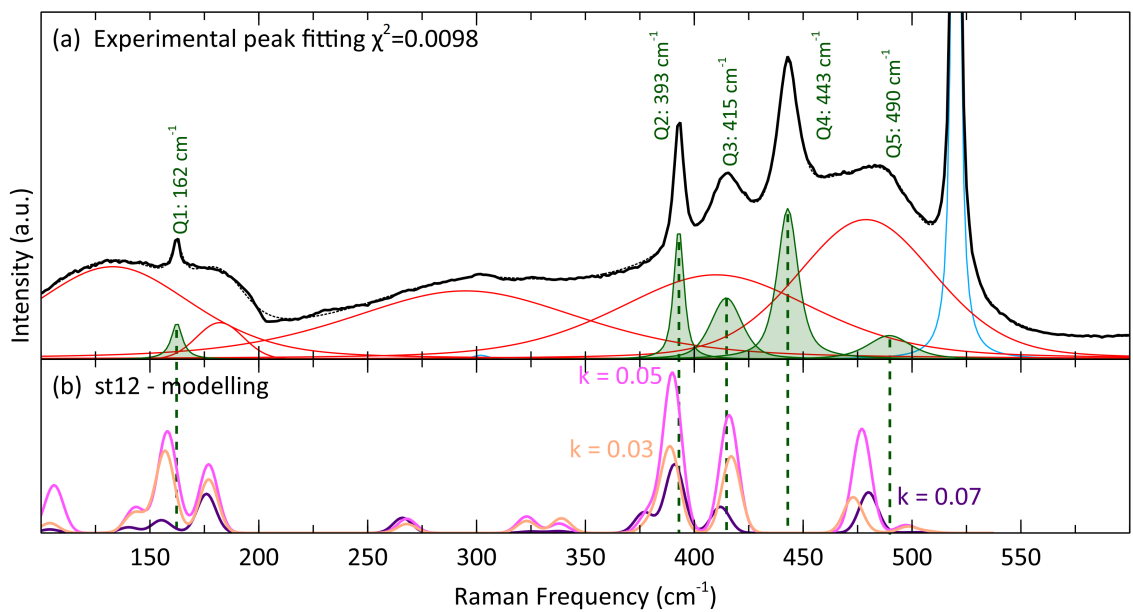

*Fig. 5. A further example of a Raman spectrum collected under confocal conditions from modifications created (a) with a 310 nJ laser pulse, which contains, in addition to a-Si (see red curve fits) and dc-Si (see blue curve fits), peaks from: compressed dc-Si and additional crystalline metastable phases (labelled as Q1 through Q5). The peak positions in (a) are given in Table 1. Also shown are the results for a computed Raman spectrum in (b) corresponding to st12-Si.*

In addition to the crystalline phases described above, there were also many instances where two additional peaks located approximately at 495 cm$^{-1}$ and 517 cm$^{-1}$ were observed. Examples are illustrated and labelled in Fig. 6. These peaks overlap with both the a-Si TO band and the dc-Si TO band and proved difficult to accurately fit in the as-recorded spectrum, Fig. 6(a). To aid the interpretation, a suitably scaled strongly a-Si spectrum, similar to Fig. 2(e), is subtracted away from the as recorded spectrum, also shown in Fig. 6(a). The resultant curve fit for the region of interest is presented in Fig. 6(b), scaled such that the two additional peaks are evident. These peaks are located at 495 and 517 cm$^{-1}$, and could be associated with hd-Si, which has been widely reported to form after annealing of r8/bc8 and exhibits similarly located Raman peaks [45]. However, the 9R polytype (stacking ABCBCACAB...) has been shown to have a similar Raman signature [46]. Furthermore, TEM selected area diffraction from this region shown in Fig. 6(c) shows a characteristic pattern that has previously been reported for the 9R polytype, with additional diffraction spots at 1/3 increments [46].

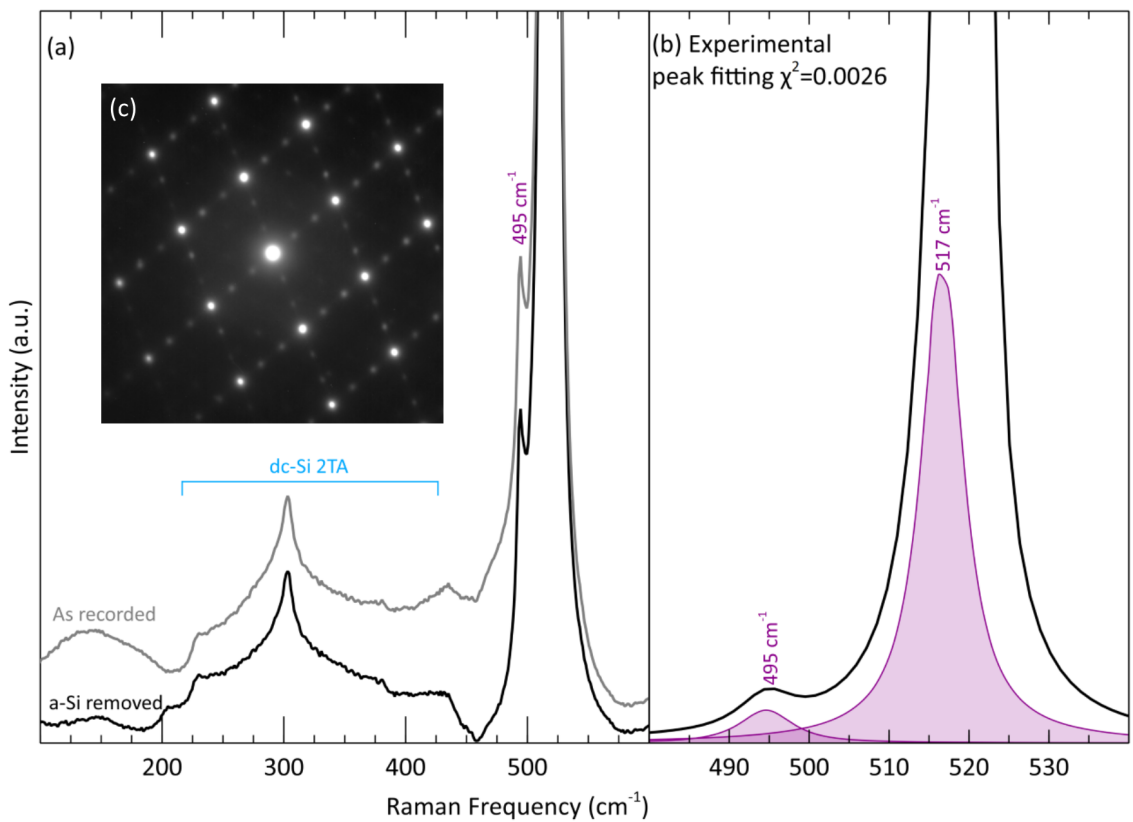

*Fig. 6. Raman spectra (a-b) collected under confocal conditions from a modification created with a 420 nJ laser pulse which contains two additional peaks just below the dc-Si TO band. The entire spectrum, (a), is shown as recorded, in grey, and in black with the a-Si component removed by subtraction of an appropriately scaled a-Si rich spectrum. In (b) the region of interest is shown with a-Si subtracted (solid black line), and the individual curves for the additional fitted peaks in purple. A selected area diffraction pattern from a TEM study of these laser-modified sites is shown in (c). It shows a characteristic 011 zone axis diffraction pattern of Si which additionally contains diffraction spots at 1/3 increments of the 111 spots which are characteristic of the previously reported 9R polytype [46].*

## IV. DISCUSSION

In the introduction it was stated that the high-intensity ultrashort laser pulses induce a dense plasma state in which crystallinity is lost. The small size of the modified sites subsequently allows for rapid quenching, particularly along the periphery at the low-intensity wings of the Gaussian spatial shape of the laser pulses. It follows that the disordered state may be preserved in the ensuing solid due to the fast cooling time via electronic heat conduction. At low fluence this time is shorter than the time required for a crystal to grow, thus resulting in the observed a-Si [3]. As fluence increases, so too does the modification size and hence the cooling time. As such, crystal growth within the modified region becomes increasingly favoured under higher fluence processing conditions. This then partly explains the varying manifestation of the a-Si in Fig. 3, where, as laser fluence increases, so too does the crystallinity. This also explains why crystalline structures are present in many samples, but not at the lowest fluence sample.

Returning to the a-Si distribution in Fig. 3, it can be seen that the variation in the a-Si intensity across modifications in the maps not only depends on the laser fluence, but also on the modification dimensions and the spatial resolution of the Raman spectrometer. In Fig. 3(a) the a-Si appears as spots since the modified regions are not spatially resolved. In Fig. 3(b), the void at the centre of the modification is large enough to be resolved, resulting in an apparent ring of a-Si. Finally, in Fig. 3(c), the modifications are large enough that the pristine region between modifications cannot be fully resolved. Indeed, as indicated in the experimental section, the limited spatial resolution of the Raman system coupled with its relatively poor detection sensitivity has other consequences for small nanocrystalline phase fractions present within a matrix of dc-Si and a-Si: in some cases they may not be detected at all by Raman. Presumably, it is for this reason that Raman peaks other than dc-Si and a-Si observed in this work appear only sporadically. They may only be observable when present in sufficiently large phase fractions and of a crystal size such that they can be observed above sensitivity limits.

Having considered why some features may be observed only sporadically with Raman, the actual nature of these features will now be discussed. In order, the remainder of the discussion will consider the compressed dc-Si, the phase assignment of the Fig. 4(a) peaks, the novel signature(s) in Fig. 5(a), and finally the peaks just below the dc-Si TO band.

Given that modifications are created by a microexplosion with a compressive shockwave, it is no surprise that evidence of compression is observable in Raman spectroscopy. The ability to maintain a remnant compressive stress in the dc-Si on the order of 4 GPa is, however, significant. Assuming similar remnant stress within the other Si phases, this raises the prospect of the existence of highly compressed a-Si (that is, high density a-Si [47,48]) and the possibility of preserved r8-Si without bc8-Si, as is also consistent with recent indentation work where the r8-Si phase is dominant in the recovered material [49]. It is also conceivable that some of the novel phases created in the far from equilibrium conditions of this work also require large remnant stresses to be preserved in the sample after laser modification. In addition, large residual stress will clearly impact on the properties (electrical and optical) of novel phases.

Treating now the crystalline phase assignment of the peaks observed in Fig. 4(a), it was noted that the peak positions in Fig. 4(a) closely resemble, but do not exactly match, the peak positions of predominantly r8 observed in the recent literature upon recovery from indentation loading [49]. They are also not a close match to the calculated peak positions for r8, bc8 and bt8-Si from the simulated spectra in Figs. 4(b), (c) and (d). Thus, in Table 2 the peak positions are shown alongside bc8/r8-Si peak positions from the literature. Firstly, past indentation reports of positions for these phases (see [50]) are in poor agreement. However, bc8/r8 measured in a DAC with a 3-3.5 GPa external pressure [51,52] is in better agreement with our experimental values particularly in the regime of 350 - 450 cm$^{-1}$. The 3 GPa values are obtained from Olijnyk & Jephcoat [51], where the ambient pressure values have been normalised to the indentation data from Table 2 and the 3 GPa peak positions shifted accordingly. The 3.5 GPa values are direct measurements from Hanfland & Syassen [52]. This agreement at residual pressure in this regime is convincing and suggests that r8-Si and bc8-Si are present. It is, however, noteworthy that the low-lying phonons in the range 160-180 cm$^{-1}$ agree less well: the laser-induced modifications do not exhibit a peak near 180 cm$^{-1}$ whereas the indentation and DAC studies under pressure do. Nevertheless, it is likely that r8/bc8-Si phases are present in our samples at residual pressures of 3.0 to 3.5 GPa.

However, the comparison between the experiment and calculations in Fig. 4, in terms of absolute peak positions of r8/bc8/bt8-Si phases, is not particularly good. As was indicated previously, the absolute peak positions and magnitudes from DFT calculations are subject to inaccuracies, particularly for very narrow bandgap or semi-metallic phases, and additionally they have been calculated at ambient pressure. Nevertheless, the calculated peaks for r8, bc8 and bt8 occur in the same wavenumber range as the experimental values (particularly in the 350-460 cm$^{-1}$ range). Thus, based on the similarity of calculated r8, bc8 and bt8 Raman signatures, it is not possible to determine unequivocally if bt8-Si is present or not from the Raman spectrum in Fig. 4(a), and indeed, to distinguish between these 3 structurally similar tetragonal phases in Raman spectra. Thus, in summary, since bt8-Si was observed in TEM cross-sections in our previous study [3], we suggest that the Raman spectrum in Fig. 4(a) may contain this phase, as well as r8 and bc8-Si, although it is impossible to determine the phase fractions of each of these phases. The fact that we observed bt8-Si but not r8/bc8-Si from TEM cross-sections previously [3] may result from the fact that thin TEM samples are essentially stress-free whereas Raman analysis of modifications in the bulk contain considerable residual stress. Direct comparison of the same modifications analysed by both Raman and TEM may resolve this issue.

*Table II. Peak positions from Fig. 4(a) as well as for bc8-Si and r8-Si from high pressure studies after unloading [50], at 3 GPa, using the data from Olijnyk and Jephcoat [51] with the ambient pressure values normalised to the indentation values in column 1 and the 3 GPa applied stress peaks shifted accordingly, and also at 3.5 GPa as observed by Hanfland and Syassen [52].*

| | Peak Position [cm$^{-1}$] | | | |
|---|---|---|---|---|
| **Peaks** | **Ambient indentation [50]** | **3 GPa [51]** | **3.5 GPa [52]** | **This work, Fig. 4(a)** |
| **P1** bc8/r8-Si | 165 | 163 | 164 | 164 |
| **P2** bc8/r8-Si | 170 | 168 | ~181 | 169 |
| **P3** r8-Si | 352 | 357 | 358 | 357 |
| **P4** bc8/r8-Si | 384 | 388 | 387 | 388 |
| **P5** bc8/r8-Si | 397 | 402 | 401 | 401 |
| **P6** bc8/r8-Si | 438 | 448 | 447 | 447 |

Now let's turn our attention to the other metastable Si signature presented in Fig. 5. In terms of the unidentified peaks in Fig. 5(a) (peaks Q1 through Q5), given that the calculated st12-Si signature (Fig. 5(b)) is a poor match, can we conclude that no known Si phase closely matches such an experimental Raman signature? In this regard, it is worth examining the accuracy of the calculated st12-Si peak positions (Fig. 5(b)) since the very narrow bandgap r8, bc8 and bt8-Si absolute peak positions were subject to significant error in DFT. However, st12-Si is expected to have a band gap greater than 1 eV [20,31] suggesting that the st12-Si Raman peak positions may be more accurate from DFT. Some further insight here can be gained by scaling the mass of known st12-Ge Raman peaks to estimate those of st12-Si based on the method of Bermejo and Cardona [53]. Indeed, using equation 1 in [54], and the experimental st12-Ge Raman data from Huston et al. [55], the estimated st12-Si peak positions were found to be in close agreement with the calculated st12-Si peak positions in Fig. 5(b). Thus, we can conclude that the experimental signature in Fig. 5(a) is unlikely to be from st12-Si. In addition, although some of the peaks in Fig. 5(a) (Table 1) somewhat coincide with r8/bc8/bt8 Raman peaks, there is again no good match to the Raman signature. Hence, the Raman spectrum in Fig. 5(a) is unlikely to contain significant phase fractions of bt8 and st12-Si and is most likely to arise from a novel Si allotrope (or allotropes) that has not previously been examined by Raman spectroscopy. This is somewhat consistent with our previous TEM study [3] that also revealed several diffraction spots that could not be indexed to known Si phases.

However, this previous TEM study [3] did observe st12-Si, so why do we not observe it in the current Raman study? There may be several reasons for this apparent inconsistency, the most likely being the fact that in our current Raman study we are observing metastable phases at significant residual pressure as indicated earlier, whereas the preparation of thin TEM cross-sections essentially releases the pressure. Hence, it is suggested that st12-Si may not be present under residual pressure above a few GPa but arises from an unknown Si allotrope at pressure (such as that giving rise to the Raman spectrum in Fig. 5(a)) that transforms to st12-Si on pressure release. Such a scenario could be examined in a future study, again by undertaking Raman and TEM from the same modification, for example, one that gives rise to the Raman spectrum in Fig. 5(a). Other factors such as very small crystal grain size, poor Raman cross-sections for these novel phases, and pressure-induced structural changes on Raman activity may also contribute to difficulty in observing particular Si phases.

In terms of identified diffractions spots in TEM and Raman spectra that cannot be correlated with known Si allotropes, DFT modelling has suggested some possibilities: namely, the monoclinic m32 and m*32 structures and the tetragonal t32 and t32* structures [3]. It would be worth exploring the structure of such phases and whether they may give Raman peaks corresponding to the Raman signature in Fig. 5(a). Indeed, correlated TEM and Raman analysis, coupled with modelling, may help identify any further novel metastable allotrope or allotropes of Si. It would additionally be desirable to pursue the creation of larger volumes of modified material, where the characterisation of the novel phases would be simplified. A novel approach of using laser pulses shaped in a Bessel-like beam would not only increase the volume of the laser-modified material by two orders of magnitude [56,57] but also has the potential to achieve higher energy density deposited into the material and thus to increase the maximum pressure in the laser-generated shock wave to the 10 TPa level [57,58].

Finally, as indicated in the results section, the additional peaks in Fig. 6(a) were located at 495 and 517 cm$^{-1}$ and could therefore be attributable to either hd-Si or 9R stacking in dc-Si. Of these two possibilities (hd-Si and 9R stacking) the former is considered unlikely for the following reasons. Firstly, hd-Si is known to arise from the thermal annealing of bc8-Si at temperatures around 200$^{\circ}$C. Indeed, after laser irradiation the fast temperature quenching of the sub-micron size modified zone indicates that the Si will essentially cool down to the solidification temperature in about 27 ns after the laser irradiation. Taking into account the speed of crystallization as 15.5 m/s [59], the 1μm Si modification is crystallised in ~65 ns, and returns to ambient temperature in about 5 μs (see supplementary of [3]). Thus, hd-Si is unlikely to arise from thermal annealing of bc8 during or following the laser-induced microexplosion. In terms of heating during Raman analysis, this only had an observable influence with both greatly prolonged exposure times, and an order of magnitude higher laser power than was used for the results presented here. Alternatively, as shown in Fig. 5(c), we have observed 9R stacking by TEM in samples exhibiting Raman peaks at 495 and 517 $cm^{-1}$. Thus, the additional peaks in Fig. 5 are much more plausibly a result of 9R stacking. Again, correlated Raman spectroscopy and TEM examination would be required to confirm this conclusion. Finally, there is the question as to how 9R stacking might arise in dc-Si in the current study and we suggest that it might be stress-induced and occur in dc-Si surrounding the highly compressed laser-modified zone. Indeed, pulsed laser melting of Si in the ns regime has also shown TEM diffraction patterns with the 1/3 spacing suggestive of 9R stacking [60].

To conclude our discussion on novel Si allotropes, we can comment on the prospect of utilising the new phases from laser-induced microexplosions for extending the current applications of Si. As indicated by others, several of the known metastable phases are likely to be narrow band gap semiconductors: r8-Si has an expected indirect band gap of 0.24 eV [20], bc8-Si has a measured (direct) band gap of 0.03 eV [21], and bt8-Si is also expected to be a narrow band gap material [3]. In addition, hd-Si is proposed to have a band gap of 0.55 eV [31], st12-Si to have a band gap between 1.1 eV [31] and 1.67 eV [20], and is additionally a candidate for superconductivity when suitably doped ($T_c$ ~35 K). Other, as yet unknown and uncharacterised phases, as in the current study and recently reported in the literature [61], will likewise be expected to exhibit similarly interesting new properties. Hence, if it were possible to tailor such phases in the near surface of Si by controlled microexplosions, then this could open up prospects for selective band gap engineering and innovative Si structures.

## V. CONCLUSIONS

This work has demonstrated that microexplosions created in a confined volume of Si by fs laser irradiation leads to a variety of Si allotropes. The main phase component observed in most modifications is a-Si, arising from rapid quenching at shorter cooling times, and this phase dominates at low laser fluences. As laser fluence increases, the modified material surrounding the void increases in volume and contains an increasing fraction of compressed crystalline Si phases. Such compression can lead to significant residual stress in modifications, measured at up to 4.5 GPa from the Raman shift of the main dc-Si peak. Since the a-Si component is normally found adjacent to this compressed crystalline region, it almost certainly exhibits a higher density than relaxed a-Si under ambient conditions. In addition, some Raman spectra at first sight appeared to reveal hd-Si but further analysis indicated that there was significant 9R stacking present and that such Raman signatures are most likely a result of this 9R polytype in dc-Si.

Raman micro-spectroscopy has also revealed several crystalline Si allotropes in addition to compressed dc-Si. There is evidence for the structurally similar phases r8, bc8 and bt8-Si, but it is impossible to determine the phase fractions of these phases from Raman data. The r8 and bc8-Si crystalline allotropes have previously been observed in near-equilibrium DAC and indentation experiments, as well as bt8 in previous microexplosion experiments, and all these phases have interesting electrical properties of relevance for electronic applications. Despite the fact that st12-Si was observed by electron diffraction spectra in our previous laser-induced microexplosion study [3], and that the simulated Raman spectra have some similarity with the experimental data in general (but do not match in terms of details), we cannot reliably assert that we observe st12-Si in the current Raman study. We attributed this inconsistency to the fact that Raman analysis of modifications in the bulk indicated compressive stresses of up to several GPa, whereas Si cross-sections thinned by FIB for TEM analysis may be under relieved compressive stress. This raises the issue that residual stress may be important in stabilizing some metastable Si allotropes, which could be important in prospective applications. In addition, both our Raman analysis in this study and the prior TEM study [3] indicated the presence of unknown Si allotropes. Clearly, in order to identify such phases, there is need for correlated TEM, Raman and XRD studies of modifications, along with DFT modelling [3].

## ACKNOWLEDGEMENTS

The authors acknowledge the support by the Australian Government through the Australian Research Council's Discovery scheme, project DP170100131. The authors acknowledge the facilities and the scientific and technical assistance of Microscopy Australia at the Advanced Imaging Precinct, Australian National University. BH acknowledges funding through ORNL's Neutron Facilities, a DOE Office of Science User Facility operated by the Oak Ridge National Laboratory. C.J.P. is supported by the Royal Society through a Royal Society Wolfson Research Merit award.